\documentclass{article}
\PassOptionsToPackage{numbers, compress}{natbib}

\usepackage[final]{neurips_2021}

\usepackage[utf8]{inputenc} 
\usepackage[T1]{fontenc}    
\usepackage{hyperref}       
\usepackage{url}            
\usepackage{booktabs}       
\usepackage{amsfonts}       
\usepackage{nicefrac}       
\usepackage{microtype}      
\usepackage{xcolor}         
\usepackage{cite}
\usepackage{aas_macros}
\usepackage{amsmath}
\usepackage{float}
\usepackage{graphicx}
\usepackage{subcaption}

\title{A Convolutional Autoencoder-Based Pipeline for Anomaly Detection and Classification of Periodic Variables}

\author{Ho-Sang, Chan\\
  Department of Physics\\
  The Chinese University of Hong Kong\\
  Sha Tin, NT, Hong Kong \\
  Center of Computational Astrophysics \\
  Flatiron Institute \\
  New York City, NY, USA \\  
  \texttt{hschan@phy.cuhk.edu.hk} \\
  \And
  Siu-Hei, Cheung \\
  Department of Physics \\
  The Chinese University of Hong Kong \\
  Sha Tin, NT, Hong Kong \\
  Center of Computational Astrophysics \\
  Flatiron Institute \\
  New York City, NY, USA \\
  \texttt{shcheungpeter@link.cuhk.edu.hk}\\
   \And 
   V. Ashley Villar \\
   Department of Astronomy \& Astrophysics \\
 Institute for Computational \& Data Sciences\\
Institute for Gravitation and the Cosmos\\
   The Pennsylvania State University \\ 
   University Park, PA, USA \\
   \texttt{vav5084@psu.edu}
  \And 
  Shirley Ho \\
  Center of Computational Astrophysics \\
  Flatiron Institute \\
  New York City, NY, USA \\
  \texttt{shirleyho@flatironinstitute.org}\\
}

\begin{document}
\maketitle
\begin{abstract}
The periodic pulsations of stars teach us about their underlying physical process. We present a convolutional autoencoder-based pipeline as an automatic approach to search for out-of-distribution anomalous periodic variables within The Zwicky Transient Facility Catalog of Periodic Variable Stars (ZTF CPVS). We use an isolation forest to rank each periodic variable by its anomaly score. Our overall most anomalous events have a unique physical origin: they are mostly highly variable and irregular evolved stars. Multiwavelength data suggest that they are most likely Red Giant or Asymptotic Giant Branch stars concentrated in the Milky Way galactic disk. Furthermore, we show how the learned latent features can be used for the classification of periodic variables through a hierarchical random forest. This novel semi-supervised approach allows astronomers to identify the most anomalous events within a given physical class, significantly increasing the potential for scientific discovery.
\end{abstract}

\section{Introduction}
Anomaly detection is an essential aspect of discovery in astronomy, as historically highlighted in the discovery of dark matter \citep{1970ApJ...159..379R}, Type Iax supernovae \citep{Li_2001}, and galaxies lacking dark matter \citep{vanDokkum2018}. As deep-sky surveys via modern telescopes continue to exponentially increase our discovery rates of variable cosmic phenomena, researchers are turning towards automated methods of anomaly detection \citep{Henrion2013, 10.1093/mnras/stz2362, 2021arXiv210312102V}.  

The time for advanced techniques to search for anomalous astrophysical events is ripe. The Legacy Survey of Space and Time (LSST) conducted by the Vera Rubin Observatory is expected to commence in 2024 \citep{2019eeu..confE..23G} and over 40 billion objects are to be observed within the 10 years of operation \citep{Ivezi__2019}. It is reasonable to expect \textit{anomalous} periodic variables stars (PVSs) which will defy expectations. Indeed, their discoveries have already been challenging our understanding of the chemical composition of stars \citep{10.1093/mnras/stab2065, Niemczura2017} and stellar evolution models \citep{2017A&A...604A..29G}. Previous data-driven studies of PVSs that utilize machine learning have focused on classifications \citep{Jamal_2020, 10.1093/mnras/stab1248, 2018NatAs...2..151N} and parameter estimation \citep{2020arXiv200507773M}. Here, we provide an anomaly detection pipeline to effectively search for peculiar PVSs detected with The Zwicky Transient Facility, a pathfinder survey for LSST. The Zwicky Transient Facility contains numerous publicly available data which serve as a good training set for big data problems in astronomy which will arise with LSST. We anticipate that our pipeline can be applied to these future surveys.

\section{Method}
\begin{figure*}[ht!] 
	\centering
	\includegraphics[width=1.0\linewidth]{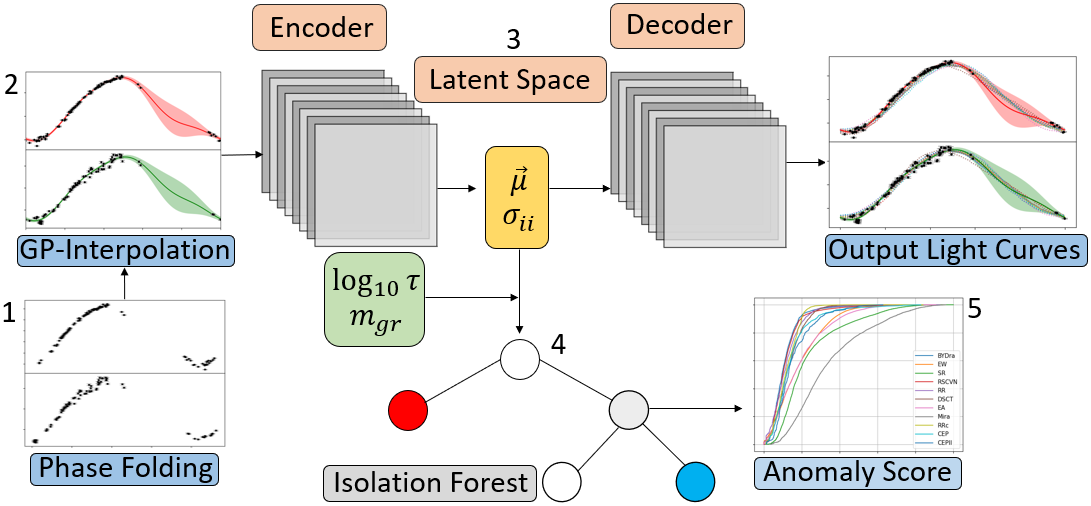}
	\caption{The anomaly detection pipeline: 1. Phase-folding the raw detection data. 2. Interpolation using the Multivariate Gaussian Process. 3. Encoding to learn latent vectors $\vec{\mu}$ and matrix elements $\sigma_{ii}$. 4. Append log$_{10}\tau$ and $m_{\text{gr}}$ to $\vec{\mu}$ 5. Isolation forest and ranking the anomalies. \label{fig:pipeline}}
\end{figure*}

Here we describe our training set and methodology. Our training set consists of the ZTF CPVS presented in \citet{Chen_2020}. The ZTF CPVS utilizes the Data Release 2 archive of the Zwicky Transient Facility \citep{2019PASP..131a8002B} to search for and classify new PVSs down to a $r$-band magnitude, a measurement of brightness, of $\sim 20.6$. They find a total of $781,602$ PVSs, of which $621,702$ are newly classified. The data are given as a multivariate, irregular time series of magnitudes, a measurement of brightness, in two filters ($g-$band and $r-$band) which are observed asynchronously. This time series is referred to as a light curve. The ZTF CPVS provides periods, which we use to phase-fold the light curves. To phase-fold a light curve, we cut the $g$- and $r$-band light curves into multiple sub-series with a time duration equal to the joint period, and we stack these sub-series on top of each other. Because the data are irregularly sampled by the telescope and taken in both bands, we choose to interpolate the phase-folded light curves using the multivariate Gaussian process \citep[MGPR,][]{dan_foreman_mackey_2014_11989, 10.1093/mnras/stz2362, 2021arXiv210604370Q} with periodic boundary conditions. The MGPR has a mean function $\eta(\phi,\lambda)$ and a covariance function $K$. We set $\eta(\phi,\lambda) = 0$, where $\phi$ is the temporal phase and $\lambda$ is a wavelength in scaled units. We choose the following covariance function:
\begin{equation}
    K(\vec{r}, \vec{r}') = C\text{exp}\Big(-\frac{|\vec{\phi} - \vec{\phi}'|^{2}}{l_{\phi}^{2}}\Big)\text{exp}\Big(-\frac{|\vec{\lambda} - \vec{\lambda}'|^{2}}{l_{\lambda}^{2}}\Big) + \begin{cases} 
      \delta & \text{if $\vec{r} = \vec{r}'$}, \\
      0 & \text{otherwise} 
\end{cases}
\end{equation}
Here, $\vec{r} = (\vec{\phi}, \vec{\lambda})$ is a high dimensional position vector, $l_{\phi}$ and $l_{\lambda}$ measure the correlation along the phase and wavelength direction respectively, $C$ is a constant, and $\delta$ measures the white noise level of the raw data. After fitting for the kernel function, we generate $160$ evenly spaced data points along with the phase direction for both $g-$ and $r-$bands. Following the approach of \citet{Villar_2020} and \citet{2021arXiv210312102V}, we stack both of them horizontally to form an `image' of size $2 \times 160$. The `images' will be encoded through a convolutional variational autoencoder \citep[C-VAE,][]{2013arXiv1312.6114K} with a LeNet structure\citep{10.1162/neco.1989.1.4.541}. Our data size is $730,184$ and we split them into a train to validation to test ratio of $7:2:1$. The architecture of the encoder is described as follows:
\begin{enumerate}
    \item \textbf{Input layer} of size $2\times 160$\vspace*{-1mm}
    \item \textbf{3 Convolutional layers} with the \textit{ReLu activation} with \textit{Dropout}\vspace*{-1mm}
    \item \textbf{Dense layer} with $256$ neurons, \textit{Linear Activation}\vspace*{-1mm}
    \item \textbf{Latent space} of size $2\times 10$
\end{enumerate}
We adopt a kernel size of $(3, 4)$ and a stride of $(1, 2)$ for the convolutional layer. The filter size of the convolutional layer increases from $32 \rightarrow 64 \rightarrow 128$, and the dropout fraction is set to $0.1$. The use of the ReLu activation is inspired by its robustness in various applications. We optimized the above-mentioned hyper-parameters through some rough grid searches, except for the kernel size and stride which are fixed. Given the variational nature of the autoencoder, the bottleneck latent space is described by a mean vector $\vec{\mu}$ and the diagonal covariance matrix $\sigma_{ii}$. Inspired by the work of \citet{10.1093/mnras/stab1248}, we apply periodic padding to the `images' during the convolution to enforce periodic boundary conditions. Our decoder is the symmetric counterpart of the encoder; however, no periodic padding and no dropout are applied. We train with the ADAM optimizer with default learning rates and a batch size of $1024$. We train for $895$ epochs for roughly $\sim6$ hours by operating on the NVIDIA RTX2080-Ti GPU. 

Once trained, we next use an isolation forest to rank each variable by a calculated anomaly score. The isolation forest works by building an ensemble of binary decision trees, which work to isolate samples of the population. The anomalies require few trees to isolate the event \citep{4781136}; in this work, we focus on events with the top 100 most anomalous scores. We use the latent vectors $\vec{\mu}$, the log of period log$_{10}\tau$, and the difference between the average $g-$ and $r-$band magnitude ($m_{\text{gr}} := \langle m_{\text{g}} \rangle - \langle m_{\text{r}} \rangle$) as input features of the isolation forest \citep{10.1145/2133360.2133363}. We note that the period and magnitudes are explicitly included because this information is lost in the pre-processing used to train our autoencoder; however, we believe that they will be valuable in filtering out anomalies. We use the isolation forest implemented in the \texttt{scikit-learn} package with based estimators of $100,000$, which we found sufficient to yield converged scores for the top anomalies. 

We additionally use our learned latent space (and some hand-engineered features, including the joint period, and the amplitude and mean magnitudes of both the $g$- and $r$- band light curves) to \textit{classify} the ZTF CPVS by their physical origin. We note that the ZTF CPVS provides class labels for PVSs, but the classifications are based on hand-engineered features only. Here we provide a new classification method that utilizes a hierarchical random forest classifier. We extracted events labels from the SIMBAD catalog \citep{2000A&AS..143....9W} by cross-matching (using \texttt{Astroquery} \citep{2019AJ....157...98G}) their sky-coordinates with those listed in the ZTF CPVS. The SIMBAD catalog contains class labels obtained, typically, through spectroscopic analysis, which is more reliable but often expensive. We found $31,541$ successfully cross-matched objects. We then construct 13 classes in 2 levels. The first level includes Active Galactic Nuclei-like (AGNL), Cepheid (CEP), Eclipsing Binaries (EB), Long-Period Variables (LPV), Mira variables (Mira), other Pulsating Variables (Pul$_\mathrm{oth}$), RR Lyrae (RR), and Peculiar (Pec). The second level is further classifications of the Pec type, and it includes Carbon stars (C-Type), Horizontal Branch stars (HB), Red Giant Branch stars (RGB), S-Type stars (S-Type), Young Stellar Object-like (YSOL), and Other Variables (V$_\mathrm{oth}$). They will serve as the data set for our classification model. We split the data set into a training-to-test set ratio of $7:3$ by using the python package \texttt{scikit-learn} \citep{scikit-learn}. We note that our training set is highly imbalanced, with the largest set containing $10,745$ events and the smallest containing just $41$ events. We balanced the training set using the python package \texttt{imbalanced-learn} \citep{JMLR:v18:16-365} with default learning parameters, which performs synthetic minority resampling \citep{chawla2002smote, JMLR:v18:16-365}. Finally, we train the hierarchical random forest classifier provided by \texttt{imbalanced-learn}, with no hyper-parameter optimization conducted.

\section{Results and Discussion}
\begin{figure*}[ht!]
\centering
\begin{subfigure}[t]{0.5\textwidth}
    \includegraphics[width=\textwidth]{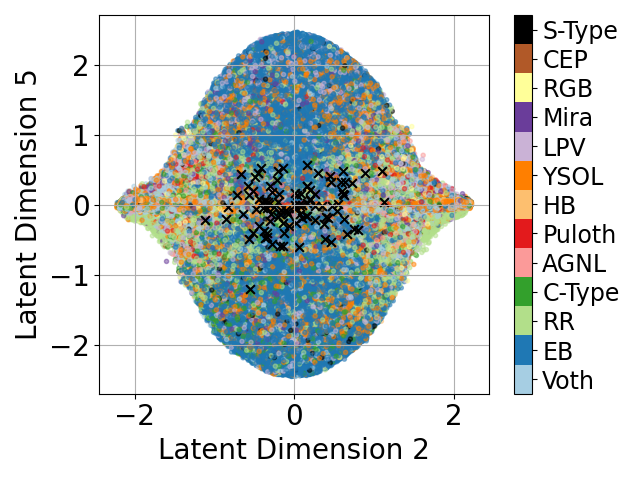}
    \caption{}\label{subfig:a}
\end{subfigure}%
\begin{subfigure}[t]{0.5\textwidth}
    \includegraphics[width=\textwidth]{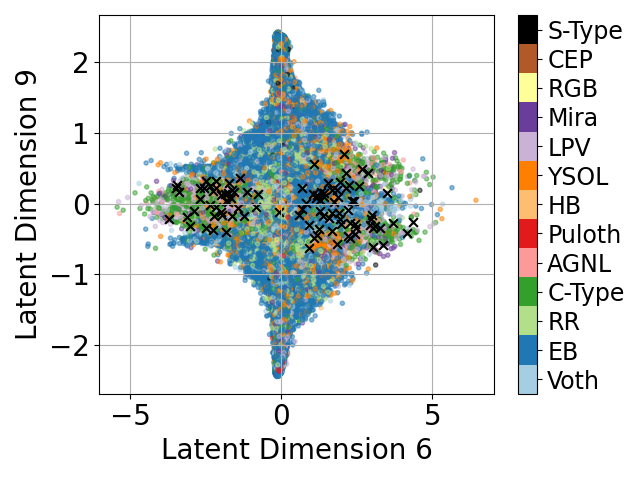}
    \caption{}\label{subfig:a}
\end{subfigure}%
\caption{(a) and (b) Examples of the latent distributions for different PVSs labeled by distinct colors. Anomalies are marked as black crosses. \label{fig:latent}}
\end{figure*}
\begin{figure*}[ht!]
\centering
\begin{subfigure}[t]{0.5\textwidth}
    \includegraphics[width=\textwidth]{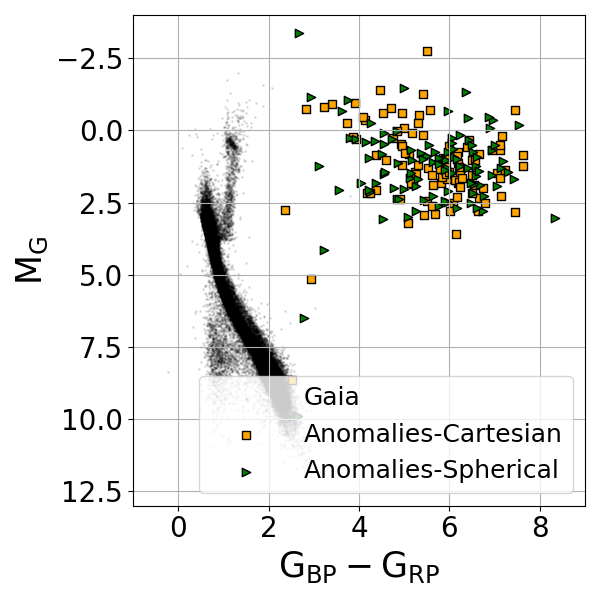}
    \caption{}\label{subfig:a}
\end{subfigure}%
\begin{subfigure}[t]{0.5\textwidth}
    \includegraphics[width=\textwidth]{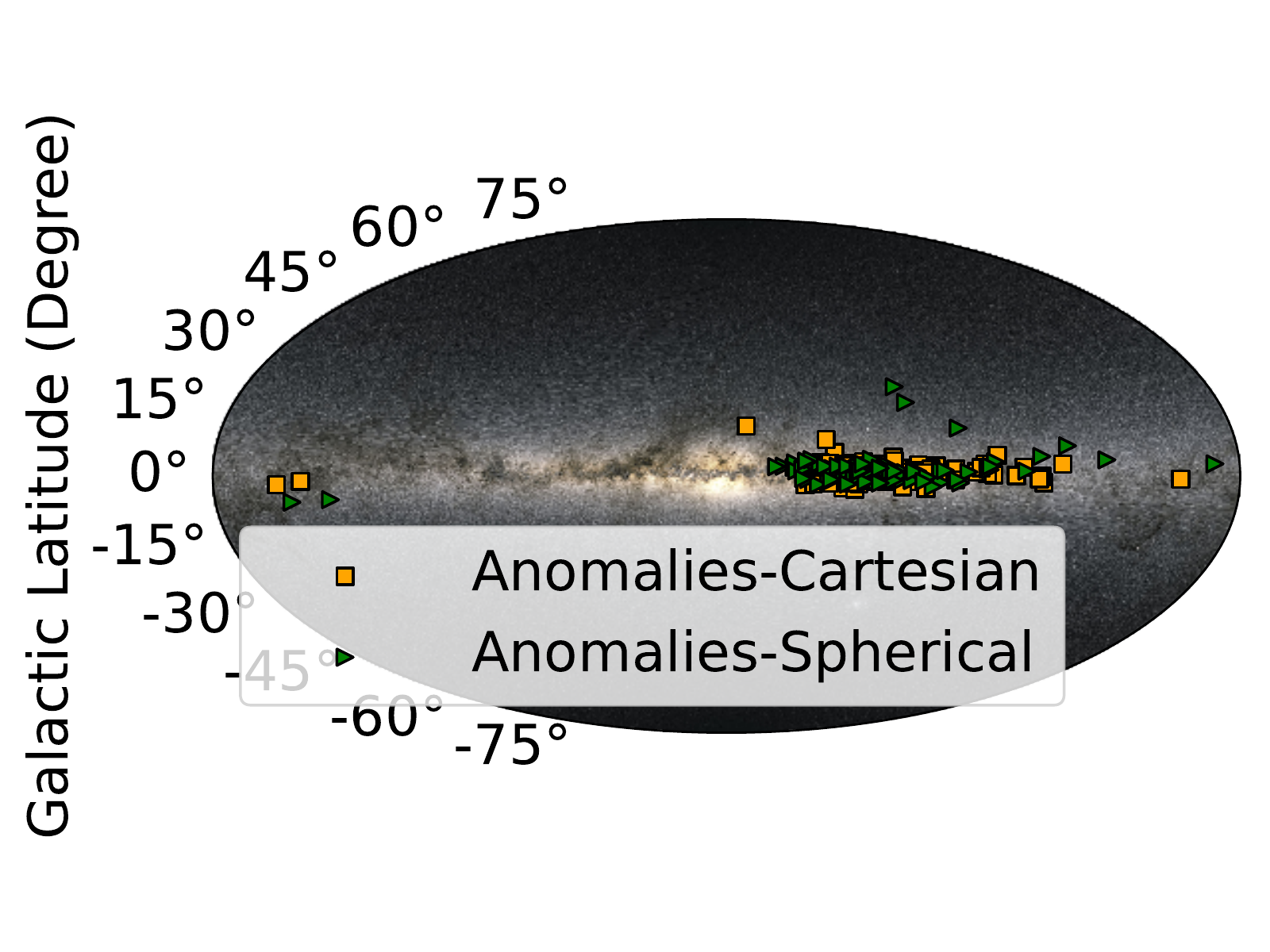}
    \caption{}\label{subfig:b}
\end{subfigure}%
\caption{(a) Distribution of the top 100 anomalous events on the Gaia HR diagram (a common feature space in astronomy) against the main-sequence stars in black dots. Data are taken from \citet{2018A&A...616A...2L}, \citet{2021AJ....161..147B}, and \citet{2021A&A...649A...1G}, through the Vizier Catalogue \citep{2014yCat....1.2023S} by \texttt{Astroquery} \citep{2019AJ....157...98G}. (b) The galactic distribution of the same set of anomalies. Note the tight distribution along the Milky Way galactic plane. Image credit: ESA/Gaia/DPAC. \label{fig:anomalies}}
\end{figure*}
\begin{figure*}[ht!]
\centering
\begin{subfigure}[t]{0.5\textwidth}
    \includegraphics[width=\textwidth]{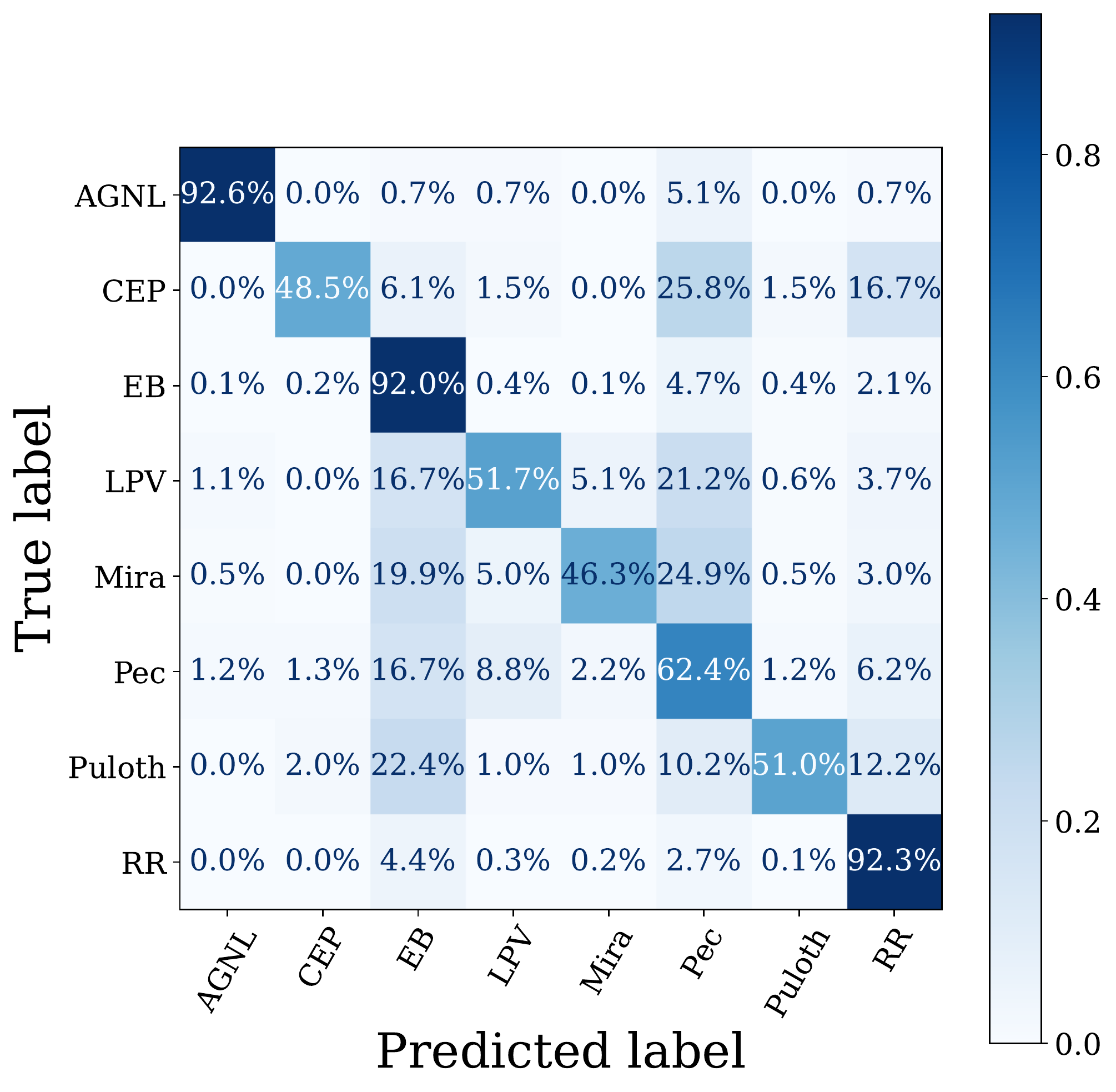}
    \caption{}\label{subfig:b}
\end{subfigure}%
\begin{subfigure}[t]{0.5\textwidth}
    \includegraphics[width=\textwidth]{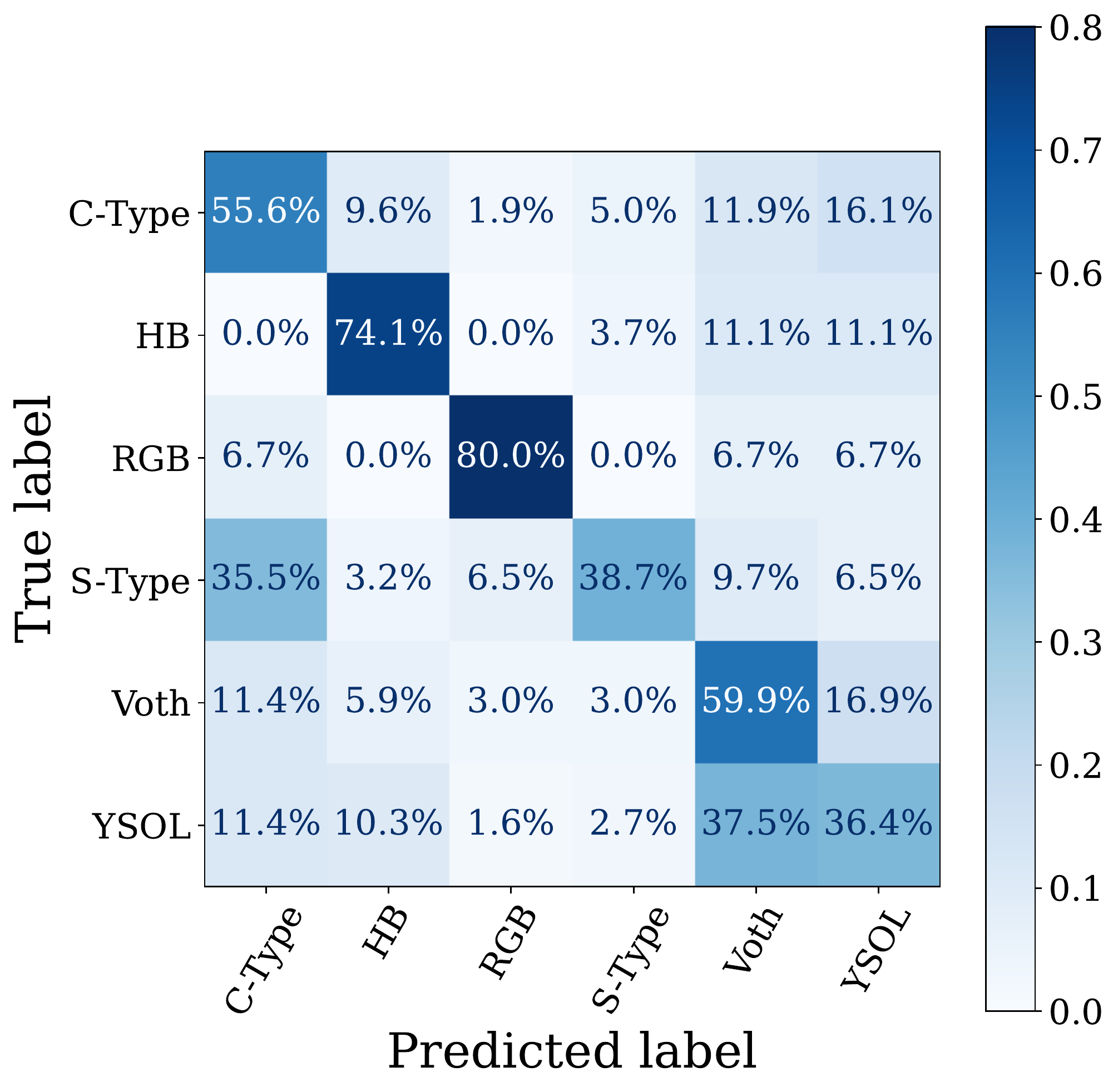}
    \caption{}\label{subfig:b}
\end{subfigure}%
\caption{(a) Confusion matrix of the first level classification labels of the test set. We show in each row the completeness of each class. (b) Same as (a), but for the second level labels (of Peculiar type). \label{fig:classbreak}} 
\end{figure*}

We find that our learned latent space exhibits an annular structure, which inspires us to transform the latent space using spherical coordinates. We run our isolation forest on both sets of coordinates and compare results, selecting the top 100 most anomalous events in both cases. Our anomaly detection algorithm is seemingly sensitive to irregular oscillations which consist of several dominant Fourier modes. In general, these anomalous events are both multi-modal and highly irregular, with some examples exhibiting larger fluctuations that span over several magnitudes. Some of the anomalies also show weak or even anti-correlation between the $g-$ and $r-$ band light curves, suggesting significant temperature variations within the pulsations. 

To better understand the nature of the selected anomalies, we extracted Gaia \citep{2021A&A...649A...1G} G-band absolute magnitudes $M_{\text{G}}$, and the difference between the Gaia B-band and R-band absolute magnitudes $G_{\text{BP}} - G_{\text{RP}}$ for the top 100 anomalies in both latent spaces. We plot their distribution in Figure \ref{fig:anomalies} (a). We note that this is a common diagnostic phase space, which roughly correlates with the temperature and luminosity of the stars. The majority of the anomalies are \textit{cool}, with $G_{\text{BP}} - G_{\text{RP}} > 4$ and \textit{luminous}, with $0 < M_{\text{G}} < 2.5$. These properties correspond to evolved, luminous, and cold stars. Furthermore, we show the distribution of the top anomalies in the Milky Way galactic coordinates in Figure \ref{fig:anomalies} (b). The majority of the anomalies concentrate in the Milky Way galactic disk, implying that (1) there is likely significant interstellar reddening due to the dust for these events and (2) the progenitor systems of these events are consistent with young and massive stars. Taken together, the observational evidence points to highly anomalous, young, cool, and massive Red Giant or Asymptotic Giant Branch stars. Spectroscopic follow-up observations and detailed light curve modeling are \textit{essential} in fully understanding the anomalies detected in our data-driven pipeline; however, we note that these anomalies were discovered with \textit{limited} survey observations. Similar techniques will be invaluable for future missions.

Finally, we highlight the classification results from our hierarchical classifier in Figure \ref{fig:classbreak} (a) and (b). We find the ZTF CPVS likely contains variable, non-stellar objects. For example, we find that $11837$ out of $97137$ ($12.2$ \%) semi-regular \textit{galactic} variable stars in the ZTF CPVS are classified as AGNL objects (objects associated with supermassive black holes in other galaxies) in our new classification model. We note that we do not use the same class breakdown as the original ZTF CPVS, making it difficult to directly use the ZTF CPVS as a baseline. A detailed comparison between classifiers will be left to future work. However, we find that our learned latent space is sufficient to classify the set of ZTF CPVS light curves explored here with reasonable levels of accuracy. We plot the latent distribution using our new labels in Figures \ref{fig:latent} (a) and (b), to show the robustness of our autoencoder in distinguishing objects that belongs to different categories.

\section{Conclusion And Broader Impact}
We present a convolutional autoencoder-based pipeline as an automatic yet robust approach to search for and classify anomalous PVSs. We note that \citet{2021MNRAS.502.5147M} recently presented an anomaly search within the Zwicky Transient Facility transient data using a deep-generative learning approach, but without a specific focus on PVSs. Our pipeline is specifically for PVSs, and could be applied to PVS data obtained from deep-sky surveys in the future. Detailed spectroscopic follow-up is essential to reveal their true identity of the anomalous PVSs identified here and how these anomalies fit into our current understanding of late-stage stellar processes.

Astrophysical science has entered the big-data era. We anticipate our method can contribute to the community, such as detecting other anomalous periodic/non-periodic transient or better classifying astronomical objects. Additionally, our unsupervised learning pipeline can make use of supervised object labels to look for categorical-wise anomalies. Techniques similar to those presented here can be used in broader applications to identify anomalies in periodic/non-periodic, multi-variate time series.

\newpage
\section{Acknowledgement}
We thank the Flatiron institute for providing computer cluster access. We thank Prof. Ming-Chung, Chu in the Chinese University of Hong Kong for providing the NVIDIA RTX2080-Ti GPU as the computational resources for the neural network training. We also thank Prof. Maria Drout and Anna O'Grady at the University of Toronto; Matteo Cantiello, Mathieu Renzo, and Adam Jermyn at the Center of Computational Astrophysics, Flatiron Institute for their valuable discussion on the anomalous periodic variables stars. VAV acknowledges support by the Simons Foundation through a Simons Junior Fellowship (\#718240) during the early phases of this project, as well as support in part by the NSF through grant AST-2108676.

\bibliographystyle{apsrev4-2}
\bibliography{main.bib}{}

\end{document}